# Peculiarities of phase states in $N_2O$–$CO_2$ cryoalloys according to electron diffraction data

A. A. Solodovnik, O. P. Konotop

B. Verkin Institute for Low Temperature Physics and Engineering of the National Academy of Sciences of Ukraine

47 Nauky Ave., Kharkiv, 61103, Ukraine

e-mail: *solodovnik@ilt.kharkov.ua*

The structural characteristics of $N_2O$–$CO_2$ alloys have been studied using transmission electron diffraction (THEED) across the entire range of their mutual concentrations at temperatures of 5 K and 65 K. The ranges in which regular solutions exist and intervals of alloy phase separation have been identified using the dependence of lattice parameters on composition and profile analysis of the diffraction pattern intensity distribution. It was found that maximum solubility values are up to 11 mol.% $CO_2$ in the $N_2O$ matrix and up to 30 mol.% $N_2O$ in solid $CO_2$ at 65 K. The solubility limits for the components at 5 K are less 20 mol.% $CO_2$ in crystalline $N_2O$ and less than 35 mol.% $N_2O$ in a $CO_2$ crystal lattice. It was established that the character of $N_2O$–$CO_2$ alloy separation corresponds to the mechanism of spinodal decomposition. The relative excess volumes that pertain to the admixture were determined for the $CO_2$ impurity in solid $N_2O$ and for $N_2O$ impurity in $CO_2$ crystal. Conclusions regarding the character of $N_2O$–$CO_2$ phase diagram have been made.

## 1. Introduction

Solid nitrous oxide ($N_2O$) and carbon dioxide ($CO_2$) belong to the class of molecular cryocrystals and exhibit similar molecular and crystalline parameters [1]. Solid molecular diluted systems are attractive objects for investigating fundamental physical phenomena, such as lattice dynamics, phase transitions, glassy states, and phase diagrams. Nitrous oxide and carbon dioxide are components of the Earth's atmosphere that make a major contribution to climate change processes [2]. Studying the condensation mechanisms in gaseous phases of $CO_2$, $N_2O$, and their mixtures is important in the context of global warming. The wide occurrence of carbon dioxide and nitrous oxide within the Solar System also attracts considerable astrophysical interest [3].

The $N_2O$ molecule differs from $CO_2$ in its symmetry and the presence of a dipole moment. According to structural studies [1], under equilibrium conditions, both crystals possess a cubic lattice, with the molecular axes oriented along the spatial diagonals of the cube (symmetry group *Pa3*). In the case of nitrous oxide, long-range "head-tail" molecular order is absent, and *Pa3* symmetry is observed only on average. According to the estimates in [4], molecular dipole ordering may occur at temperatures below 11 K (space group $P2_13$). Monte Carlo simulations [5] predict a stable *Pa3* symmetry group within the temperature range of 0–300 K. It should be noted that the realization of "head-tail" order is theoretically possible even in the absence of dipole–dipole interactions. For this phenomenon, it is sufficient that the "quasi-center" of the molecule does not coincide with a lattice site. The space group $P2_13$, calculated for solid $N_2O$ in [4], has not been observed experimentally in structural studies [6–8]. The problem of the possible existence of local regions exhibiting ordered "head-tail" alignment of $N_2O$ molecules has been discussed in several works. Indirectly, the asymmetry of the $N_2O$ molecule may manifest itself in certain physical characteristics; thus, minor anomalies in the thermal [9] and optical [10] properties of $N_2O$ crystals have been attributed to the formation of small clusters with "head-tail" order. However, heat transport studies of solid $N_2O$ have revealed additional phonon scattering relative to $CO_2$ crystals due to “head-tail” disorder” within the crystal lattice [11, 12]. The mechanisms of realization of *$P2_13$* and *Pa3* space symmetry groups have been investigated through head-tail flip calculations in [13], where it was shown that no short-range dipole order exists in solid $N_2O$.

Interest in solid mixtures of nitrous oxide and carbon dioxide is also driven by the unique properties of solid $N_2O$. An unexpected effect of spontaneous dipole ordering in $N_2O$ cryocondensates was identified in [14]. The generation of an electric potential between the substrate and the sample surface accompanied this phenomenon. The structures of condensates at 62 K and 40 K exhibited distinct characteristics. To explain these results, a theoretical model [15, 16] was proposed, based on the competition between dipole arrangement and thermal effects. The findings established that the spontaneous-electric effect is a novel phenomenon in solid-state physics, and nitrogen oxide is a representative of a new class of spontaneous-electric materials.

The ability of nitrous oxide to form isomorphous states is of particular interest. Studies of the interaction of condensed $N_2O$ films with solid surfaces revealed an isomorphic state in which the $N_2O$ molecule can realize two canonical structures: $N^{-}{=}N^{+}{=}O$ and $N{\equiv}N^{+}{-}O^{-}$ [15]. These $N_2O$ isomers differ in the character of their valence bonds, the outer-shell electronic configuration, and the magnitude of their dipole moments [2, 17]. Quantum-mechanical calculations reported in [16] showed that, in addition to the ground state of the N–N–O molecule, there are two further local energy minima – a relatively stable cyclic isomer and an unstable linear N–O–N isomer. Utilizing the concept of isomorphous states, the authors of [19, 20] explained the ability of $N_2O$ samples to generate "cryoradiation" during the condensation process. This effect manifested as discrete, high-intensity flashes visible to the naked eye. The proposed model is based on the existence of several possible molecular isomers. It was hypothesized that the formation of non-equilibrium radiation states results from the presence of $N_2O$ isomers in the gas phase and their subsequent thermally stimulated transformation. According to the hypothesis, flashes represent a superposition of a series of relaxation processes of individual regions connected to each other. Significant evidence for the existence of $N_2O$ isomers is the luminescent emission at 516.8 nm recorded in [21], which is close to the transition energy (2.81 eV) from the cyclic isomer to the molecular ground state.

Preliminary data regarding the structure of solid $N_2O$–$CO_2$ mixtures were obtained by Vegard [22] via X-ray diffraction of samples containing 25, 50, and 75 mol.% $CO_2$; however, these data are insufficient for a complete description of the phase diagram. Previously, in the study of $N_2O$–$CO_2$ cryosolutions [23], we determined the dependences of the lattice parameter on the composition of the solutions, which exhibited a non-monotonic complex behavior; in particular, the plots revealed an abrupt step-like change in lattice-parameter values. The objective of the present study is a detailed investigation of the morphology and structure of $N_2O$–$CO_2$ cryoalloys across the entire range of mutual concentrations. Particular emphasis is placed on phase transformations in solid solutions with a predominant $N_2O$ content.

## 2. Experiments

The $N_2O$–$CO_2$ alloys were investigated by transmission high-energy electron diffraction (THEED) method using a cryostat integrated with an EG-100A electron diffractometer. The experiments were performed within a temperature range of 5–65 K. The electron diffraction technique is promising for detecting new phases, since it permits the identification of a newly formed phase when domain sizes correspond to several tens of angstroms. The experimental procedure and sample preparation were described in detail previously [24]. Particular aspects directly relevant to the present study are noted below. The composition of the gas mixture was determined by the difference in partial pressures, with the concentration measurement error not exceeding 0.1 %. Samples were prepared by deposition of the gas mixture onto a combined aluminum–carbon substrate. The crystalline Al film served as an internal standard, providing the required accuracy for lattice-parameter determination; the resulting error did not exceed 0.005 Å. The amorphous carbon substrate was utilized to measure the integrated intensity of diffraction reflections, with an associated error of 6–7 %. Substrate surfaces were chemically neutral, preventing interaction between the sample surface and the substrate. As a result, no surface-induced changes in the state of the $N_2O$–$CO_2$ condensates were observed. In solid solutions predominantly composed of $N_2O$, the emergence of the spontaneous dipole ordering effect is hindered by $CO_2$ impurities, which modify the anisotropic intermolecular interaction. A distinctive advantage of the electron-diffraction method is the ability to visually observe the processes in real time on a screen inside the diffractometer column. The generation of cryoradiation [20], typically characterized by discrete flashes, was neither visually observed nor recorded on the electron diffraction patterns. The realization of this phenomenon is prevented by the vacuum conditions in the cryostat, as the effect does not occur below a pressure threshold. The employed experimental methodology ensured the formation of radiation-equilibrium states during the condensation of the gas mixture. Necessary minimization of the electron beam's radiative impact was achieved by use of a Faraday cylinder, which blocked the electron flux during temperature stabilization. Diffraction patterns were recorded photographically, then scanned and processed digitally.

## 3. Results and discussion

In $N_2O$–$CO_2$, the initial components satisfy the Hume-Rothery requirements for unlimited solubility [25], as well as Kitaigorodsky's solubility conditions for molecular crystals [26]. These conditions require similarity in particle shape and size, as well as matching structural symmetry – the structures of the mixed substances must be isomorphous. Both nitrous oxide and carbon dioxide crystals exhibit the same lattice type: face-centered cubic (fcc). Theoretically, these factors should ensure the formation of a continuous series of solid solutions. However, the dependence of the lattice parameter on concentration exhibits a nonmonotonic, complex character. Comparison of experimental data and theoretical patterns should be performed for regular solutions, since discrepancies in results arise from both the character of the components' distribution in the mixture and the degree of alloy dispersion. The study of binary solid systems imposes specific requirements on sample preparation. Regular solutions are solutions whose lattice model assumes a random distribution of particles in a cubic lattice [27]. To facilitate this randomness, the following preparation conditions were implemented. The gas mixture was prepared at room temperature to ensure homogeneity. The condensation rate was maintained below 0.3 nm/s to prevent molecular fractionation in the flow. Sample growth occurred at temperatures low enough to maintain condensation coefficients close to unity for both components, ensuring that the alloy composition matched that of the condensed gas mixture.

One of the deviations from the theory of an ideal solution is manifested through the anomalous behavior of the dependence of the lattice parameter on the composition. The general view of concentration dependence of the lattice parameters for $N_2O$–$CO_2$ alloys at 65 K (black squares) and 5 K (red circles) is presented in Fig. 1 [23]. The plots exhibit a complex nature with several distinct regions. In the low-concentration range, the dependence follows Vegard's law, indicating the formation of substitution solid solutions. We note some key peculiarities of the $N_2O$–$CO_2$ system. The main unexpected effects are observed in solid mixtures with predominant $N_2O$ content. The most significant distinction between high- and low-temperature solutions lies in the location of a "step-like" jump in lattice parameter values. At 65 K, this sharp change occurs in a narrow region near 48 mol.% $CO_2$, whereas at 5 K, the effect shifts to approximately 20 mol.% $CO_2$. Another positional deviation is also observed: at 65 K, the horizontal interval of the plot lies above the Vegard dependence; at 5 K, it falls below it. No temperature influence on the character of the lattice-parameter dependence for $CO_2$-based solid solutions has been noted. Substitutional solutions form at 65 K in samples with $N_2O$ impurity content up to 30 mol.% and up to 11 mol.% $CO_2$ in $N_2O$-based solutions. In low-temperature alloys, approximately 20 mol.% $CO_2$ dissolves in the $N_2O$ lattice, while approximately 35 mol.% of impurity dissolves in the $CO_2$ lattice. The broader range of existence of substitutional solutions within the $CO_2$ matrix is attributed to the positional uncertainty of the asymmetric $N_2O$ molecule at the cubic lattice sites and its ability to displace along the space diagonals of the cube. Conversely, the position of the symmetric $CO_2$ molecule is strictly defined, with the carbon atom always occupying the lattice sites. But since the displacements of the $N_2O$ molecules are uncorrelated, the spatial symmetry is preserved "on average" and corresponds to the *Pa3* space group.

Processes occurring within solid solutions can be analyzed using the concept of relative excess volumes per isolated impurity [28]. Determining excess volume values is essential for resolving issues related to mixing volumes and for analyzing the contribution of rotator impurities to thermal expansion. A correlation exists between excess volume values and the extent of solid solution intervals; obtaining these data provides insights into the lattice deformations surrounding impurity particles. Assuming the impurity particles are isotropic, the formula for the relative excess volume ($\Delta V/V$) per isolated impurity for cubic lattices was derived in [15] as follows:

$$\frac{\Delta V}{V} = \frac{V_{(x)} - V_{(0)}}{xV_{(0)}}, \qquad (1)$$

where x is the solution concentration and V(x) is the volume of the solution at concentration x.

The calculated values of $\Delta V/V$ at 65 K are as follows:

- $\Delta V/V = -0.06$ in an $N_2O$ matrix;
- $\Delta V/V = 0.07$ in a $CO_2$ matrix.

These results indicate that an $N_2O$ impurity molecule in a $CO_2$ matrix occupies a larger volume than a matrix molecule, with a relative magnitude of 1.07. Conversely, in an $N_2O$ matrix, the volume per impurity molecule is expected to be smaller than that of a matrix molecule, equaling 0.94.

At 5 K, the values are:

- $\Delta V/V = -0.02$ in an $N_2O$ matrix;

- $\Delta V/V = 0.07$ in a $CO_2$ matrix.

The influence of temperature on the relative excess volume per isolated impurity in alloys with a predominant $CO_2$ composition is negligible. In low-temperature solutions based on $N_2O$, the crystal lattice is loosened, and the excess volume per impurity molecule (0.98) is greater than in an ideal solution. Around ~20 mol.% $CO_2$, a sharp decrease in the lattice parameter is observed, accompanied by a change in the alloy morphology. At 65 K, the dissolution of $CO_2$ in $N_2O$-based alloys leads to crystal contraction. Strains around the impurity in the $N_2O$ lattice increase upon the addition of $CO_2$ to the mixture, raising questions regarding the stability of the solid solution. The dependencies shown in Fig. 1 exhibit a plateau – a region where lattice parameter values remain independent of composition change, which is a characteristic sign of phase decomposition. Discrepancies between the lattice parameter values and the phase composition can be interpreted as a change in morphology. Within the range of constant lattice parameter values, a system of diffraction reflections corresponding to the *Pa3* space group is observed.

It is difficult to conclude that a system is single-phase based solely on the presence or absence of diffraction reflections of its constituent components. The nature of the lattice parameter behavior in solid solutions at 65 K was investigated using full-profile analysis of the intensity distribution of the diffraction maxima. Figure 2 shows the diffractogram of the sample with a composition of 5 mol.% $CO_2$, which corresponds to the substitutional solid solution region. The observed reflections, approximated by Lorentzian functions (the red curve is an approximation of the full experimental diffraction profile and is equal to the sum of the blue curves, which correspond to discrete maxima), belong to the *Pa3* symmetry, except for a low-angle maximum whose origin will be discussed later. The diffraction maxima (110), (210), (310), which are characteristic in the case of head-tail ordering, are absent. Figure 3 presents the diffraction patterns of alloys with compositions of 25 mol.% $CO_2$ (a) and 37.8 mol% $CO_2$ (b), corresponding to the interval of presumed decomposition. The approximation confirms a single system of *Pa3* reflections. No maxima attributable to diffraction at twin boundaries were detected.

Furthermore, the absence of stacking faults (SF) is confirmed by the symmetry of the diffraction maxima. In the presence of SF, a mutual shift of the (111) and (200) reflections toward each other would typically occur. However, no such shift was recorded.

Analysis of the diffractograms showed that an increase in the dopant component is accompanied by a change in the half-width of the diffraction lines. Figure 4 illustrates the concentration dependence of the full widths at half maximum (FWHM) for the (111), (210), and (220) diffraction lines. As evident from the plot, the incorporation of $CO_2$ molecules into the $N_2O$ lattice results in a nearly linear increase in FWHM. This effect is attributed to processes occurring within the cryoalloy, since the only varying parameter is concentration. The observed broadening does not allow separation of coexisting phases, but leads to the idea that a substantial contribution to the broadening arises from small crystallites of a new phase or from a new phase with lattice parameters similar to the host phase.

Analysis of the ratio between intensities of regular and superlattice reflections provides important information. The latter are determined by the non-central interaction between molecules and depend on sample composition. The concentration dependence of the intensity ratio of the (210) to (220) maxima is presented in Figure 5. Within the substitutional solid solution region (0–11 mol.% $CO_2$), the ratio of the superlattice reflection intensity to the regular reflection intensity decreases relative to the pure component, since incorporation of impurity into the solution alters the anisotropic interaction. In the plateau region, the dependence of the (210)/(220) ratio on composition is weak and equals, on average, the value observed for samples with 25 mol.% $CO_2$, which supports the assumption of solid solution decomposition. The nature of the $N_2O$–$CO_2$ cryoalloy separation, occurring without separation into distinct components, corresponds to spinodal decomposition, which proceeds without formation of nuclei (nucleation) [29, 30]. Spinodal decomposition – one of the two possible mechanisms for the phase separation of a homogeneous solution – occurs spontaneously when a homogeneous system becomes unstable at a specific composition or temperature. As a result, phases with the same crystal lattice but different compositions and without clear boundaries between them are formed. The emerging phases are not always stable, and the decomposition process may continue. This circumstance can explain the differences in the intensities of diffraction maxima observed in Figures 3a and 3b. The separation process is determined by the nature of the interparticle interactions within the solution, leading to the formation of two or more new phases. Notably, in the molecular-atomic $N_2O$–Ar system [8], the interaction between molecules and atoms is different, and the decomposition is binodal, which occurs via a nucleation mechanism. In this case, the formation of new phase nuclei is followed by their subsequent growth. Upon addition of up to 10 mol.% Ar, the $N_2O$–Ar solid solution decomposes into its constituent components, $N_2O$ and Ar, which have different crystal structures.

Thus, the boundaries of the decomposition region for the $N_2O$–$CO_2$ cryosolution have been established as 11–70 mol.% $CO_2$. An increase in $CO_2$ content up to 48 mol.% in the mixture led to an unexpected effect: a sharp decrease in the alloy's lattice parameter within a narrow concentration range. This step-like anomaly exhibits the characteristics of a first-order phase transition. In the $CO_2$ alloy, this first-order phase transformation is realized through structural changes in the rotational subsystem of the mixture, consisting of symmetric and asymmetric molecular rotators [31]. After completion of the parameter jump, alloys based on the carbon dioxide crystal lattice are observed in the equimolar region. A distinctive feature of the 50 mol.% samples is the dependence of their morphology on temperature. At 65 K, the diffraction patterns of the alloy correspond to the *Pa3* space group structure, whereas condensation at 5 K led to the formation of an amorphous phase. Variation of the sample composition from equimolar by a few percent up or down led to the formation of a crystalline phase. In low-temperature alloys, the step anomaly occurs within a narrow concentration range around 20 mol.% $CO_2$. In this case, within the spinodal decomposition region, the phases exhibit a carbon dioxide lattice structure.

On the diffractogram of the $N_2O$–$CO_2$ solution with 5 mol.% $CO_2$ (Fig. 2), an additional maximum is present at small angles, which does not belong to the *Pa3* structure. This reflection is also observed in helium-temperature samples at low $CO_2$ concentrations. A possible explanation for the additional reflection is based on a model of isomorphic molecular states of nitrous oxide. It is known that, in addition to the classical linear structural formula, two other molecular formulas for $N_2O$ exist: cyclic and a linear form with an alternative atomic arrangement [18, 32]. The gas phase of $N_2O$ is a mixture of molecular isomers at a certain concentration that differs from that in the solid phase. This assumption, proposed by the authors of [19, 20], facilitated the interpretation of their experimental data. In the present case, it is assumed that small additions of $CO_2$ influence the formation of isomorphic molecular states within the nitrous oxide matrix and that the additional phase is associated with this effect. Further research is required for definitive conclusions.

## 4. Conclusions

The structural characteristics of $N_2O$–$CO_2$ alloys produced at temperatures of 5 K and 65 K have been investigated across the entire range of their mutual concentrations using transmission electron diffraction (THEED). Specimens were prepared by depositing the gaseous mixture on substrates. The structure of $N_2O$–$CO_2$ samples with predominant $N_2O$ or $CO_2$ belongs to the face-centred cubic lattice with *Pa3* symmetry.

To determine the solubility regions in $N_2$ –$CO_2$ alloys, the concentration dependence of the solid mixture lattice parameter and the diffraction pattern intensity distribution have been studied.

It was found that up to 20 mol.% $CO_2$ may be dissolved in the $N_2O$ matrix while the $N_2O$ solubility in solid $CO_2$ is about 35 mol.% at helium temperature. Maximum solubility values are up to 11 mol.% for $CO_2$ in $N_2O$ crystal and up to 30 mol.% for $N_2O$ in $CO_2$ lattice at 65 K.

The ranges of low fractions of both components have been studied in detail, and the relative excess volumes that pertain to the admixture are determined for the $CO_2$ impurity in solid $N_2O$ and for $N_2O$ molecules in $CO_2$ matrix.

It was established that the character of $N_2O$–$CO_2$ cryosolution separation corresponds to the mechanism of spinodal decomposition.

Solid $N_2O$ and $CO_2$ have similar physical properties that create preconditions for forming close to ideal substitutional solid solutions. However, the $N_2O$–$CO_2$ system shows a non-ideal, complicated behaviour of the alloys. This “anomaly” is related to the mixing process, which changes the intermolecular interaction and the orientational positioning of the molecules in the lattice. The role of the orientational subsystem should be taken into account. The behaviour of molecular alloys is mainly determined by the character of intermolecular interaction.

Acknowledgement.

Special thanks to E. Korshikov for stimulating discussions and for the assistance in finding the literature.

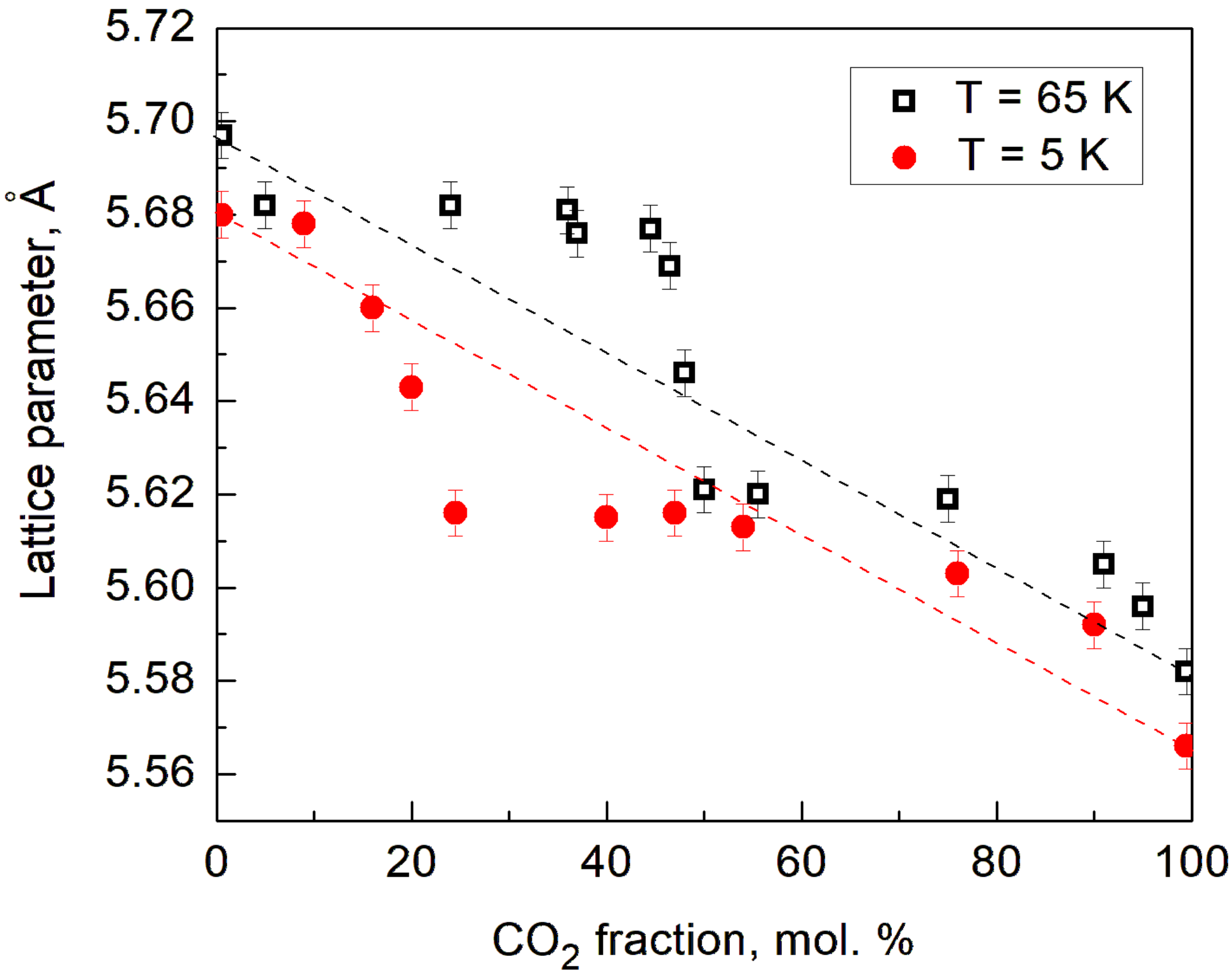

Fig. 1. Dependences of the lattice parameter of solid $N_2O$–$CO_2$ mixtures on the concentration of carbon dioxide. The condensation temperature is 5 K (red circles) and 65 K (black squares). The dashed lines correspond to Vegard's rule.

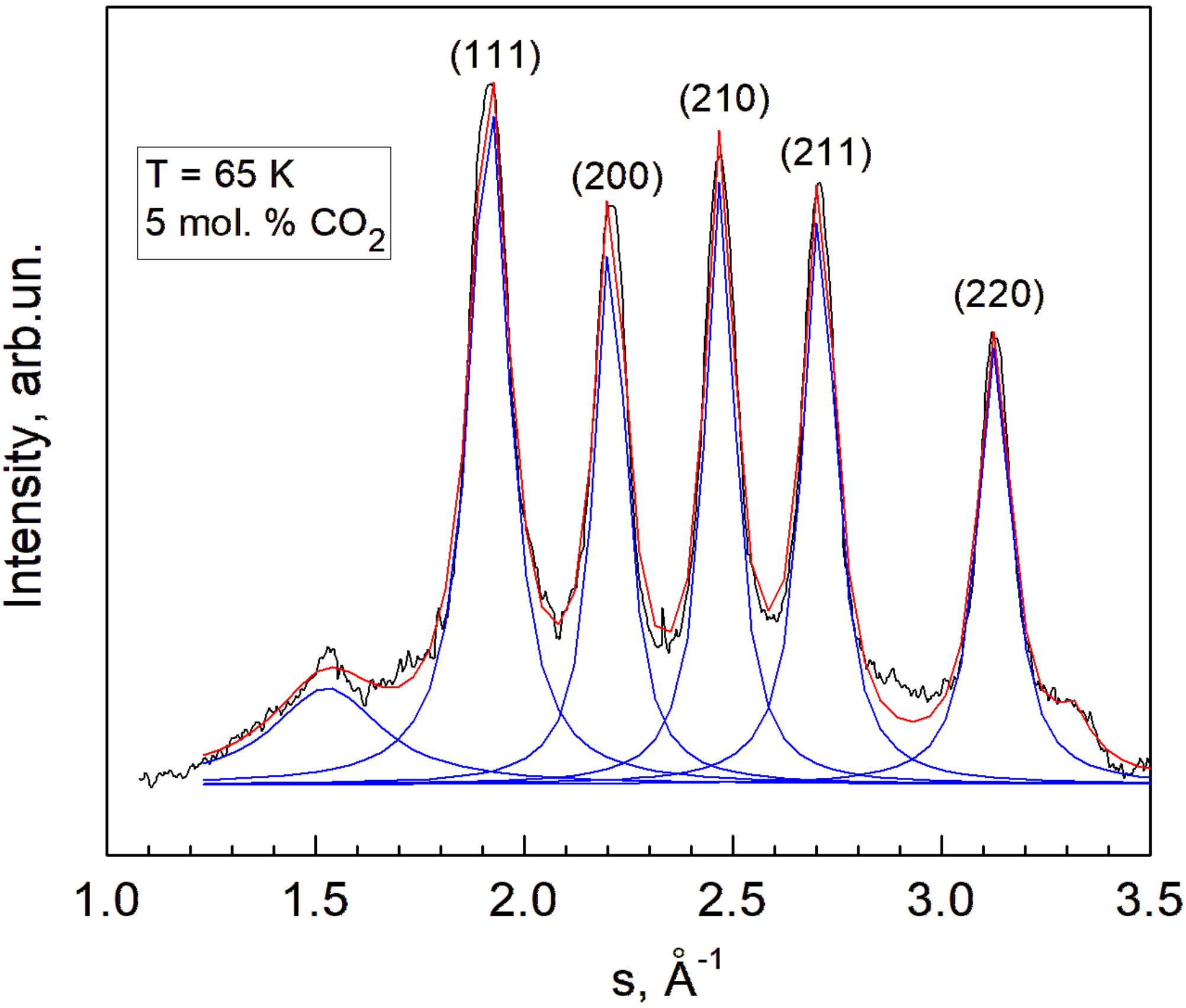

Fig. 2. Section of the diffractogram from the $N_2O$–$CO_2$ sample with a composition of 5 mol.% $CO_2$ at a deposition temperature of 65 K. Red and blue curves are approximations of experimental data by Lorentz functions.

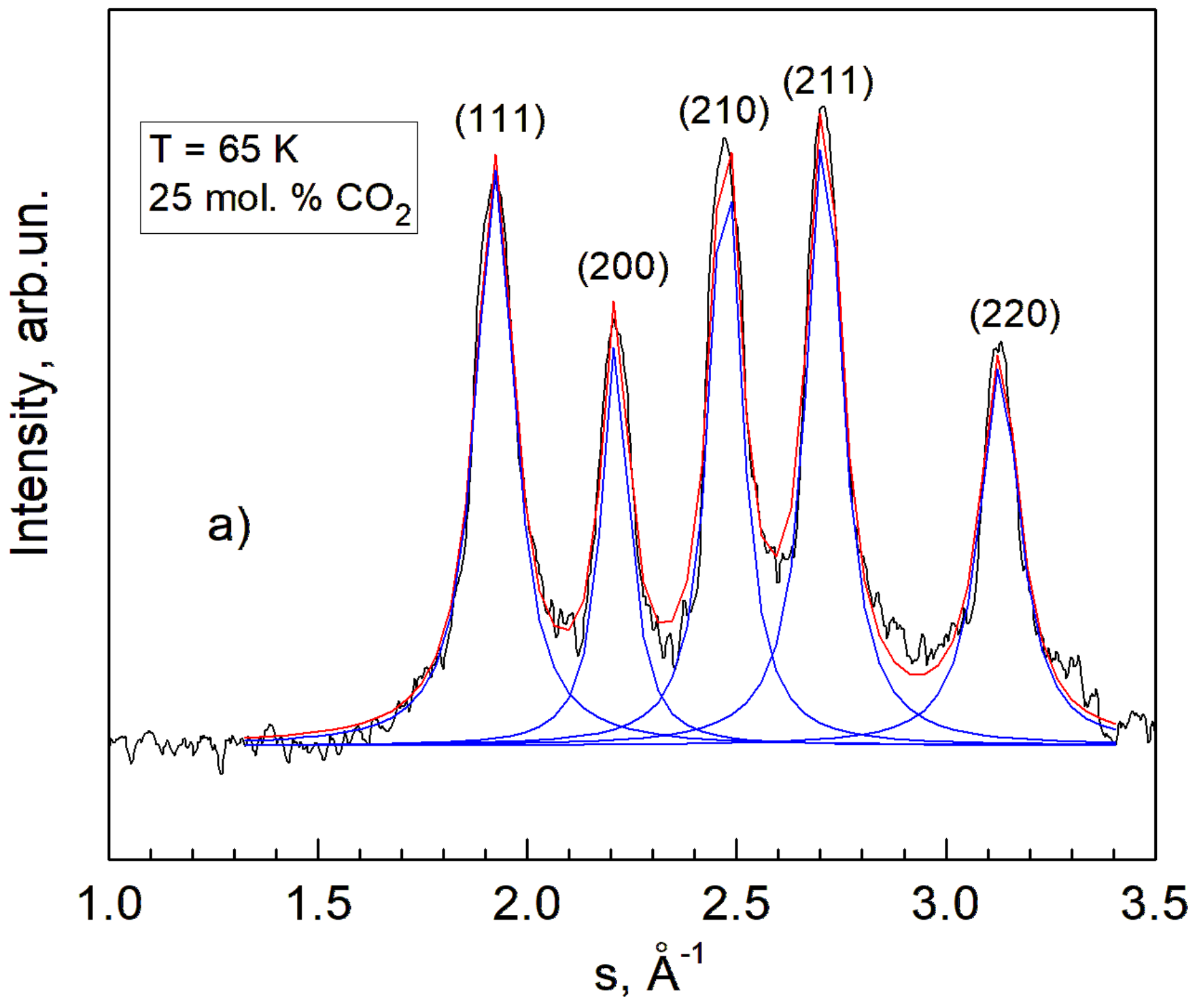
T = 65 K
25 mol. % CO2
(111)
(200)
(210)
(211)
(220)
a)
Intensity, arb.un.
1.0
1.5
2.0
2.5
3.0
3.5
s, Å-1

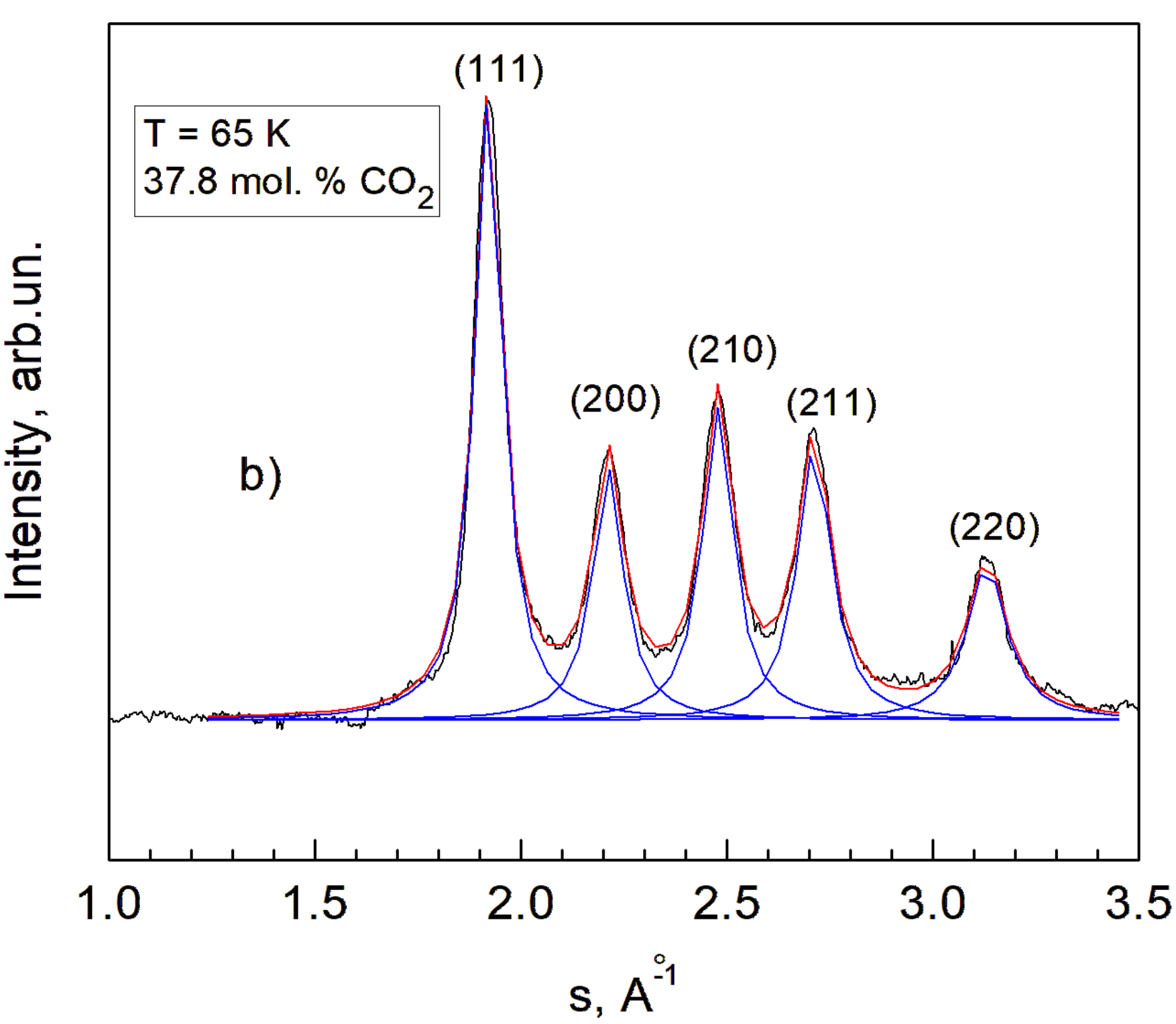
T = 65 K
37.8 mol. % CO2
(111)
(200)
(210)
(211)
(220)
b)
Intensity, arb.un.
1.0
1.5
2.0
2.5
3.0
3.5
s, Å-1

Fig. 3. Sections of diffractograms from $N_2O$–$CO_2$ samples with a composition of 25 mol.% $CO_2$ (a), 37.8 mol.% $CO_2$ (b), at a deposition temperature of 65 K. Red and blue curves are approximations of experimental data by Lorentz functions.

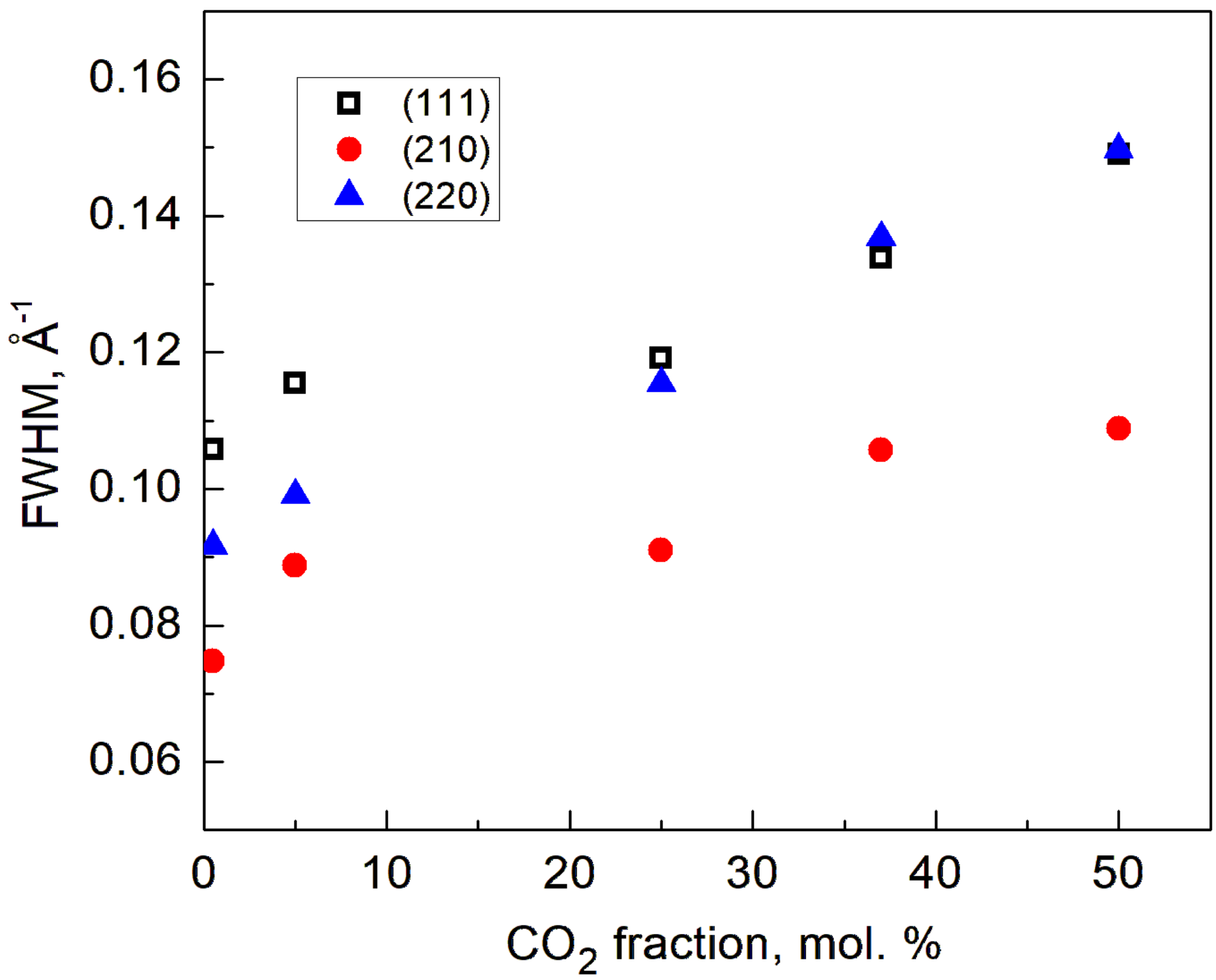

Fig. 4. Dependences of the full widths at half maximum (FWHM) of the (111), (210), and (220) diffraction maxima on the concentration of carbon dioxide for $N_2O$–$CO_2$ samples deposited at a temperature of 65 K.

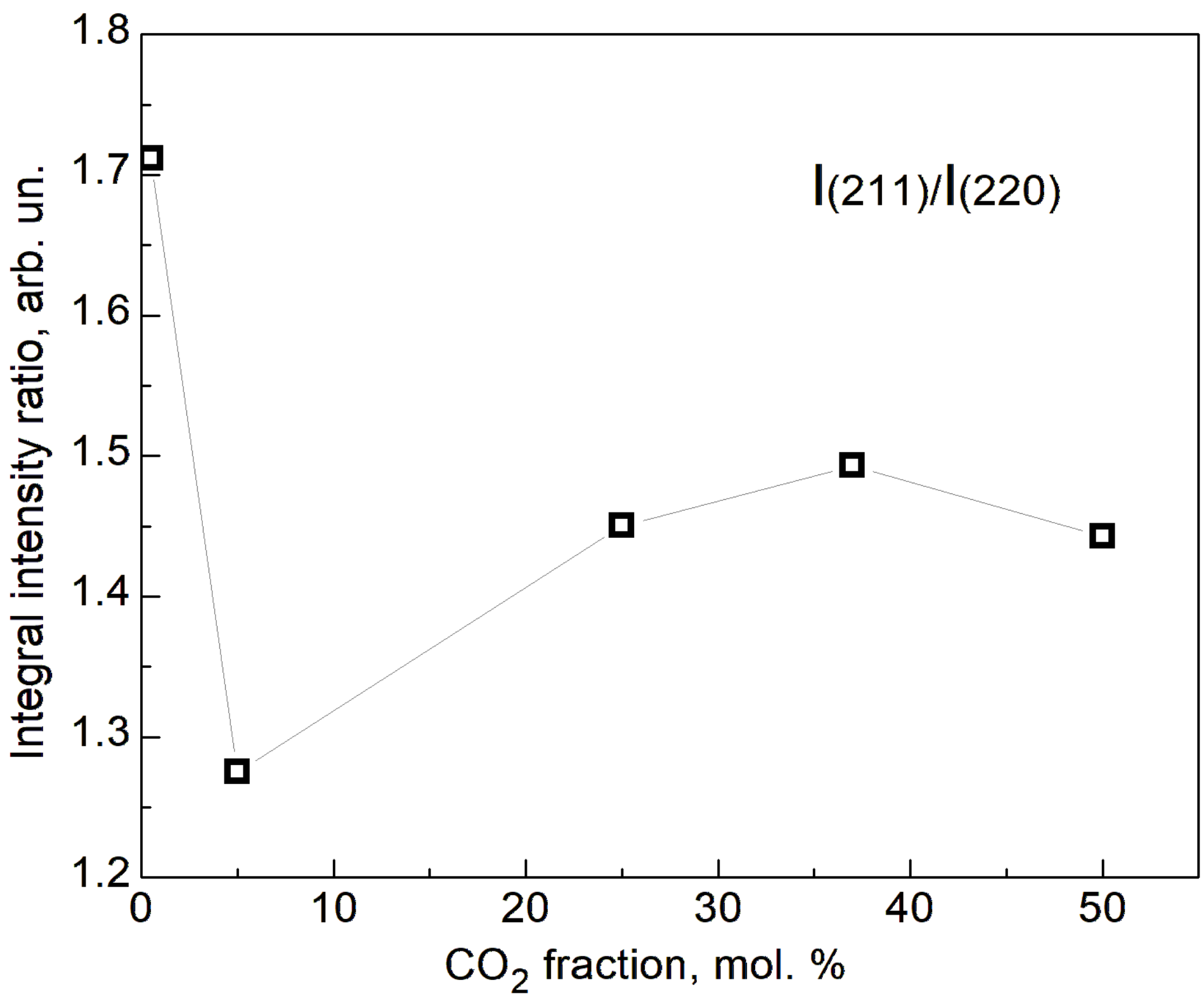

Fig. 5. Dependence of the ratio of the intensities of the (211) superlattice and (220) regular diffraction maxima on the concentration of carbon dioxide for $N_2O$–$CO_2$ samples deposited at a temperature of 65 K.

1. *Physics of Cryocrystals,* edited by V. G. Manzhelii, Yu. A. Freiman, V. L. Klein, A. A. Maradudin, AIP Press, Woodbury (1997).
2. W. C. Trogler, *Coord. Chem. Rev.* **187**, 303 (1999). https://doi.org/10.1016/S0010-8545(98)00254-9
3. R. N. Clark, R. Calson, W. Grundy and R. N. Noll, *Observed Ices in the Solar Systems*, in: *The Science of Solar System Ices*, M. S. Gudipati, J. Castillo-Rogez (eds), *Astrophysics and Space Science Library* **356**, Springer, New York (2012). https://doi.org/10.1007/978-1-4614-3076-6_1
4. M. W. Melhuish and R. L. Scott, *J. Phys. Chem.* **68**, 2301 (1964). https://doi.org/10.1021/j10079a046
5. B. Kuchta, R. D. Etters, *J. Chem. Phys*. **95**, 5399 (1990). https://doi.org/10.1063/1.461656
6. W. C. Hamilton, M. Petrie, *J. Phys. Chem.* **65**, 1453 (1961). https://doi.org/10.1021/j100826a509
7. S. I. Kovalenko, E. I. Indan, A. A. Khudoteplaya, I. N. Krupskii, *Phys. Stat. Sol. (a)* **20**, 629 (1973). https://doi.org/10.1002/pssa.2210200225
8. A. A. Solodovnik, V. V. Danchuk, *Low Temp. Phys.* **29**, 788 (2003) [*Fiz. Nizk. Temp.* **29**, 1041 (2003). https://doi.org/10.1063/1.1614193
9. A. M. Tolkachev, V. G. Manzhelii, V. P. Azarenkov, A. Jezowski, E. A. Kosobutskaya, *Sov. J. Low Temp. Phys.* **6**, 747 (1980) [*Fiz. Nizk. Temp.* **6**, 1533 (1980)]. https://doi.org/10.1063/10.0030190
10. A. A. Anderson, *Chem. Phys. Lett.* **8**, 537 (1971). https://doi.org/10.1016/0009-2614(71)80086-6
11. V. A. Konstantinov, V. G. Manzhelii, S. A. Smirnov, A. M. Tolkachev, *Sov. J. Low Temp. Phys.* **14**, 104 (1988) [*Fiz. Nizk. Temp.* **14**, 189 (1988)]. https://doi.org/10.1063/10.0031877
12. V. V. Sumarokov, P. Stachowiak, J. Mucha, A. Jezowski, *Phys. Rev*. **B74**, 224302 (2006). https://doi.org/10.1103/PhysRevB.74.224302
13. K. R. Nary, R. Kuhus, M. S. Conradi, *Phys. Rev.* **B26**, 3370 (1982). https://doi.org/10.1103/PhysRevB.26.3370

14. R. Balog, P. Cieman, N. C. Jones, D. Field, *Phys.Rev. Lett.* **102**, 073003 (2009). https://doi.org/10.1103/PhysRevLett.102.073003

15. D. Field, O. Plecan, A. Cassidy, R. Balog, N. C. Jones, J. Dunger, *Int. Rev. Phys. Chem.* **32**, 345 (2013). https://doi.org/10.1080/0144235X.2013.767109

16. A. Cassidy, O. Plecan, R. Balog, J. Dunger, D. Field, N. C. Jones, *J. Phys. Chem. A* **118**, 6615 (2014). https://doi.org/10.1021/jp503332x

17. G. Geballos, H. Wende, K. Baberschke, D. Arvantis, *Surf. Sci.* **15**, 482 (2001). https://doi.org/10.1016/50039-6028(01)00938-4

18. F. Wang, R. D. Harcourt, *J. Phys. Chem.* **104**, 1304 (2000). https://doi.org/10.1021/jp9930088

19. A. Drobyshev, A. Aldijarov, E. Korshikov, V. Kurnosov, D. Sokolov, N. Tokmolin, *Low Temp. Phys.* **39**, 460 (2013) [*Fiz. Nizk. Temp.* **39**, 591 (2013)]. https://doi.org/10.1063/1.4807327

20. A. Drobyshev, A. Aldijarov, E. Korshikov, Y. M. Strzhemechny, *Low Temp.* Phys. **41**, 547 (2015) [*Fiz. Nizk. Temp.* **41**, 703 (2015)]. http://dx./doi.org/10.1063/1.4927046

21. A. Drobyshev, Yu. Strzhemechny, A. Aldijarov, E. Korshikov, V. Kurnosov, D. Sokolov, *J. Low. Temp. Phys.* **187**, 71 (2017). https://doi.org/10.1007/s10909-016-1693-7

22. L. Vegard, *Z. Physik.* **71**, 465 (1931). https://doi.org/10.1007/BF01340080

23. A. A. Solodovnik, V. V. Danchuk, *Low Temp. Phys.* **35**, 339 (2009) [*Fiz. Nizk. Temp*. **35**, 438 (2009)]. https://doi.org/10.1063/1.3117965

24. A. A. Solodovnik and N. S. Mysko-Krutik, *Low Temp. Phys.* **45**, 545 (2019) [*Fiz. Nizk. Temp*. **45**, 637 (2019)]. https://doi.org/10.1063/1.5097365

25. W. Hume-Rothery, R. E. Smalman, C. W. Howorth, *The Structure of Metals and Alloys,* The Metals and Metallurgy Trust, London (1969).

26. A. Kitaigorodsky, *Molecular Crystals and Molecules*, Academic Press, New York (2012).

27. J. H. Hildebrand, R. L. Scott, *Regular Solutions,* Prentice-Hall, New York (1962).

28. M. A. Strzhemechny, A. A. Solodovnik, S. I. Kovalenko, *Low Temp. Phys*. **24**, 669 (1998) [*Fiz. Nizk. Temp*. **24**, (1998)]. https://doi.org/10.1063/1.593662

29. K. Binder, *Rep. Prog. Phys.* **50**, 783 (1987). https://doi.org/10.1088/0034-4885/50/7/001

30. M. Hillert, *Acta Met.* **9**, 525 (1961). https://doi.org/10.1016/0001-6160(61)90155-9

31. K. A. Chishko, A. A. Solodovnik, *Low. Temp. Phys*. **52**, 366 (2026) [*Fiz. Nizk. Temp.* **52**, 366 (2026). https://doi.org/10.1063/10.0042903

32. H. Okabe, *Photochemistry of Small Molecules*, John Wiley & Sons, New York (1978).